# The Extremes in Intra-Night Blazar Variability: The S4 0954+65 Case

Rumen Bachev *, Anton Strigachev, Evgeni Semkov, Rosa Victoria Muñoz Dimitrova, Georgi Latev, Borislav Spassov and Blagovest Petrov

Institute of Astronomy, Bulgarian Academy of Sciences, 72 Tsarigradsko Shosse Blvd, Sofia 1784, Bulgaria; anton@astro.bas.bg (A.S.); semkov@astro.bas.bg (E.S.); rdimitrova@astro.bas.bg (R.V.M.D.); glatev@astro.bas.bg (G.L.); bspassov@astro.bas.bg (B.S.); bpetrov@astro.bas.bg (B.P.) * Correspondence: bachevr@astro.bas.bg

**Abstract:** We present results of optical observations of the extremely violently variable blazar S4 0954+65 on intra-night time scales. The object showed flux changes of up to 100% within a few hours. Time delays between optical bands, color changes and "rms-flux" relations are investigated and the results are discussed in terms of existing models of blazar variability.

## 1. Introduction

S4 0954+65 is a gamma-ray loud, "superluminal"-jet object at $z$ = 0.368. Its R-band magnitudes are typically around 16, except for the beginning of 2015, when the object reached R = 13 (Bachev 2015 [1] and the references therein). Previous reports on this object indicate high activity on intra-night time scales [2–6]. In this work we study the intra-night variability of S4 0954+65 in different optical bands.

## 2. Observations

The blazar S4 0954+65 ($z$ = 0.368) was monitored on intra-night time scales with the 60-cm telescope of Belogradchik observatory, Bulgaria for a total of 64 h between February 2015 and April 2016. The telescope is equipped with an FLI PL9000 CCD and standard BVRI filter set [7]. The average monitoring duration for each run was about 3.5 h. The object was observed semi-simultaneously in several wavebands (BVRI).

## 3. Results

During the time of our monitoring that happened around its historical maximum, S4 0954+65 showed exceptional intra-night variability, very rarely seen in other objects. On some occasions, changes reached 0.5–0.7 magnitudes over several hours. Microvariations of up to 0.2 mag/h were frequently observed. No reliable time lags between the bands were detected, considering the time resolution of the datasets, which is about 5–10 min (details in Bachev 2015 [1]). Chromatic behavior was observed on longer time scales. The object clearly appears to be more active (in terms of fractional variability) during high states.

Figures 1 and 2 show the long-term behavior of the blazar in three different colors (BRI) and the color changes respectively. Figures 3–5 show examples of the observed rapid intra-night variations. Figure 6 shows the observed "rms-flux" relation, clearly indicating higher rms when brighter.

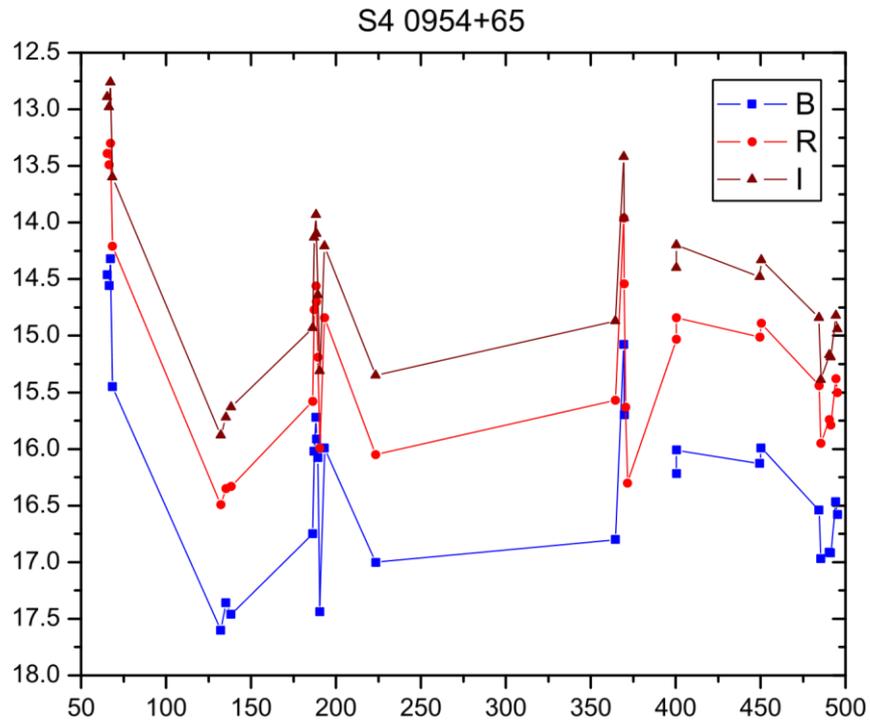

**Figure 1.** Long-term behavior of S4 0954+65 for the observed period (February 2015–April 2016). Changes of up to 3 mag are seen for a relatively short period of time.

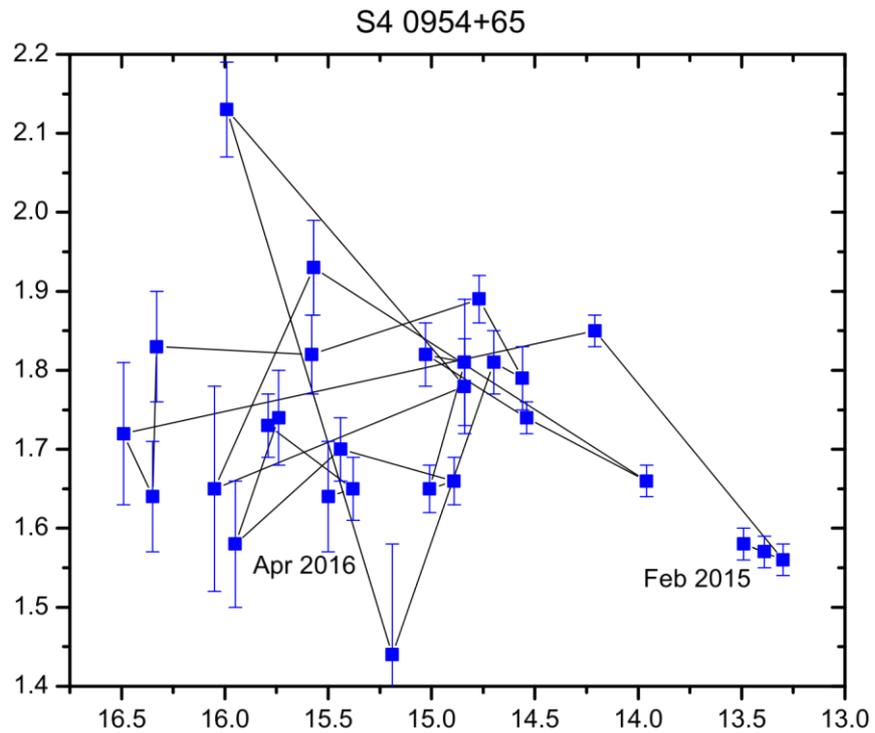

**Figure 2.** Color (B-I) index for the observed period as a function of the R-band magnitude. No clear systematic behavior can be traced, making color changes to appear random to a large extent.

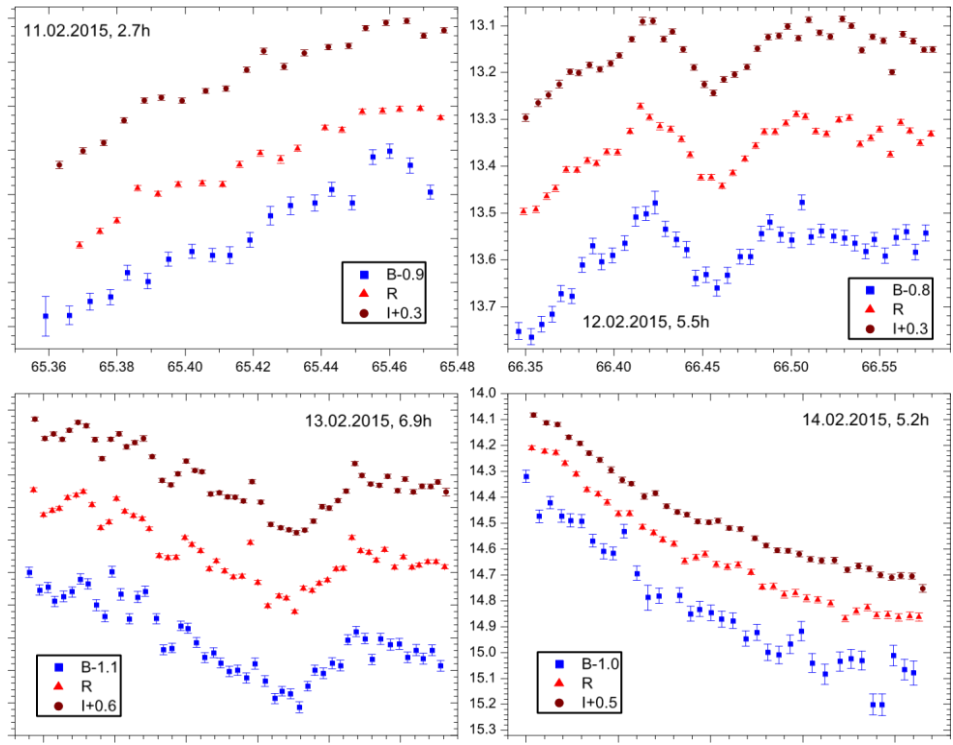

**Figure 3.** Examples of rapid intra-night variability.

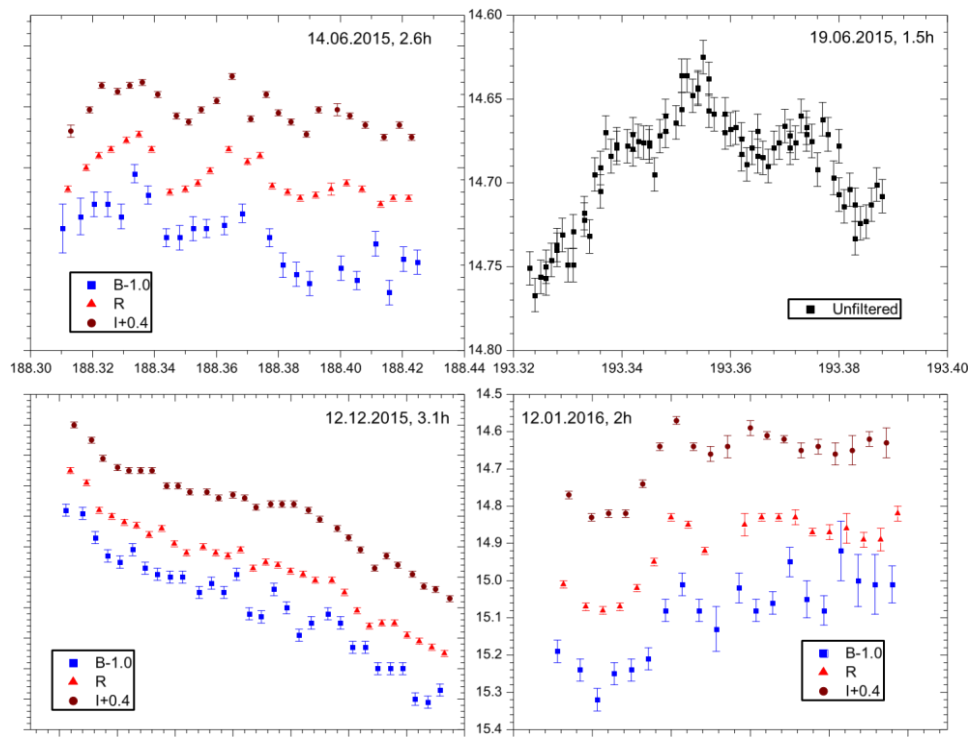

**Figure 4.** Examples of rapid intra-night variability.

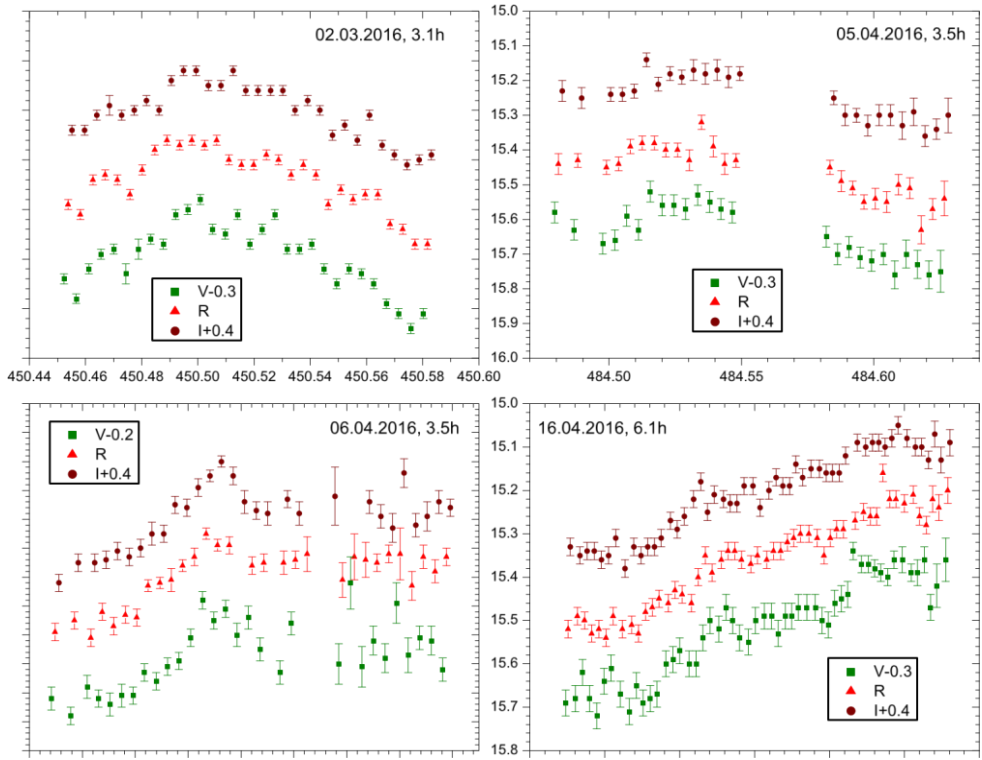

**Figure 5.** Examples of rapid intra-night variability.

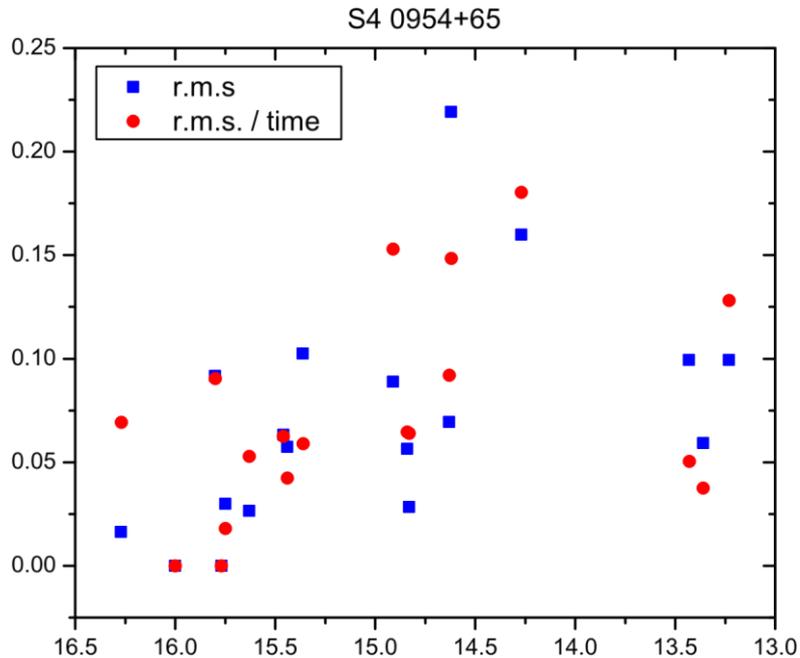

**Figure 6.** "rms-flux" relation for all intra-night observing sets. A clear tendency of higher rms in brighter states can be seen (correlation coefficient of about 0.5). Blue squares show rms, calculated for each observing set, and the red circles are the time normalized rms, to account for the possible influence of the duration of the set. Both show similar trends.

## 4. Discussion and Conclusions

The color changes on long-term scales that appear almost random perhaps imply the presence of more than one emitting component, each producing a different SED and evolving differently. Higher fractional variability during high states implies the presence of multiplicative type interaction between the emitting components (e.g., avalanche type) instead of additive type, when the "fractional variability—average flux" relation should have the opposite trend.

The "rms-flux" relation seems to favor a Doppler factor $\delta$ change (e.g., blobs travelling along a swinging jet) as responsible for the fast variability, as $F \sim \delta^{3+\alpha}$, where $\alpha$ is the spectral index, $F_\nu \sim \nu^{-\alpha}$, meaning that even a single emitting blob of large Doppler factor can account for both high average flux and large rms, while the total number of blobs may play only a minor role. In contrast, if the intra-night variability is due to fast evolution (turning on and off) of a number (N) of emitting blobs, where the individual Doppler factors change little during the time of observation, then one would expect $1/\sqrt{N}$ behavior, i.e., the opposite "rms-flux" trend should be observed.

Such and similar studies, especially on very active objects like S4 0954+65, may help to significantly better understand the physics of the relativistic jets. Our monitoring of this object is ongoing.

**Acknowledgments:** This work was partially supported by the Bulgarian NSF through grant DNTS DO 02/137.

**Author Contributions:** Rumen Bachev did some of the observations, most of the data analysis and wrote the text. Evgeni Semkov and Anton Strigachev helped with the data analysis and useful discussions and suggestions. Rosa Victoria Mu?oz Dimitrova, Georgi Latev and Blagovest Petrov did some of the observations.

**Conflicts of Interest:** The authors declare no conflict of interest.